\documentclass[12pt]{article}
\usepackage{graphicx}

\begin{document}

\title{All-spherical telescope \\ with extremely wide field of view}
\author{V.~Yu.~Terebizh$^{1,2}$\thanks{E-mail: valery@terebizh.ru}\\
$^{1}$Crimean Astrophysical Observatory, Nauchny, Crimea 298409\\
$^{2}$Institute of Astronomy RAN, Moscow 119017, Russian Federation}

\date{July 25, 2015}

\maketitle

\begin{abstract}
An all-spherical catadioptic system made of glass of one type is proposed 
for the monitoring of large sky areas. We provide an example of such 
a system with the aperture of diameter $400$~mm and the curved field 
of $30^\circ$ in diameter.
\end{abstract}
 
\begin{figure}
  \includegraphics[width=84mm]{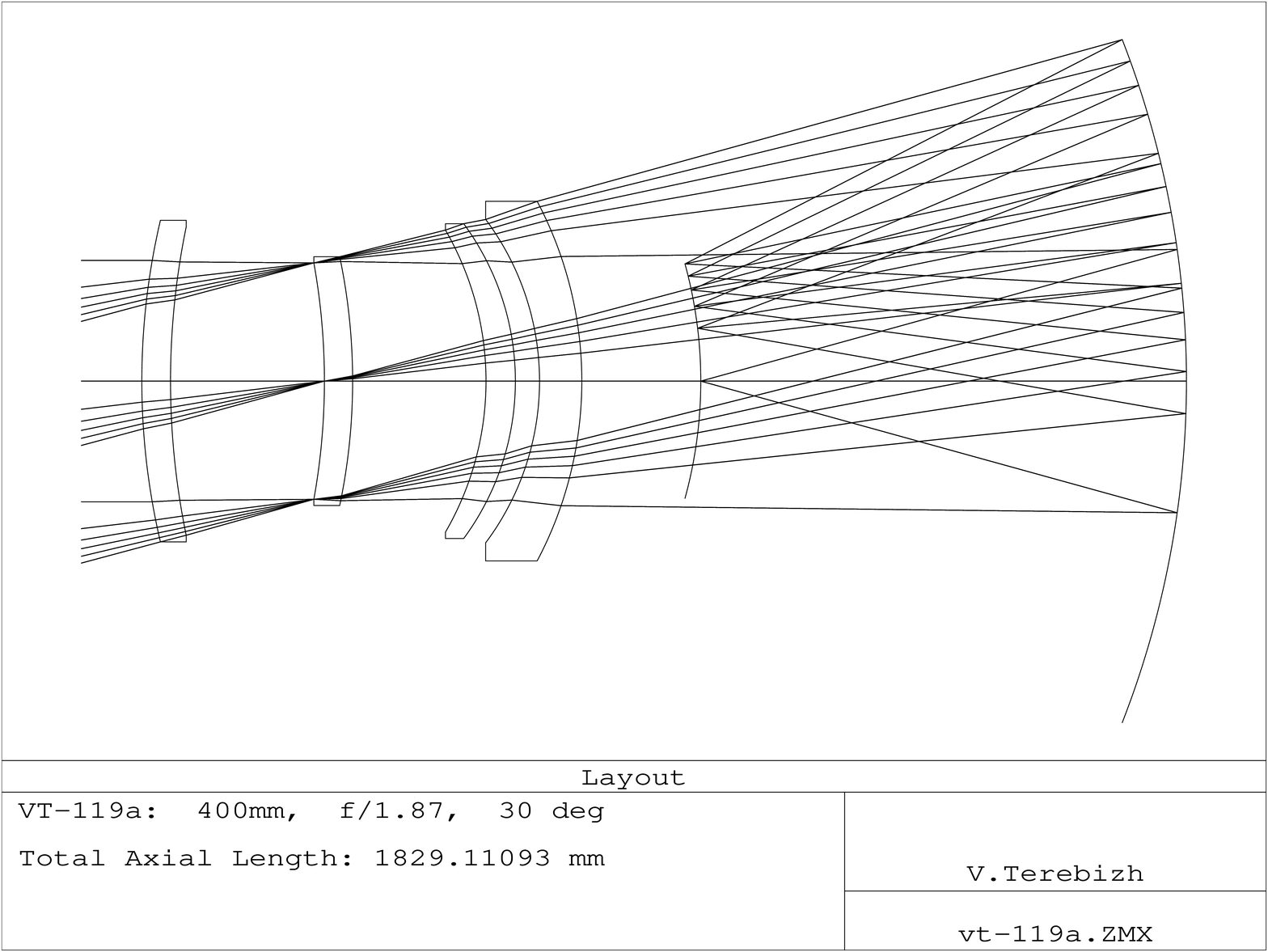} 
  \caption{Optical layout of design VT-119a with entrance pupil diameter of 
    $400$~mm and a field of $30^\circ$.}
\end{figure}

\begin{figure}
  \includegraphics[width=84mm]{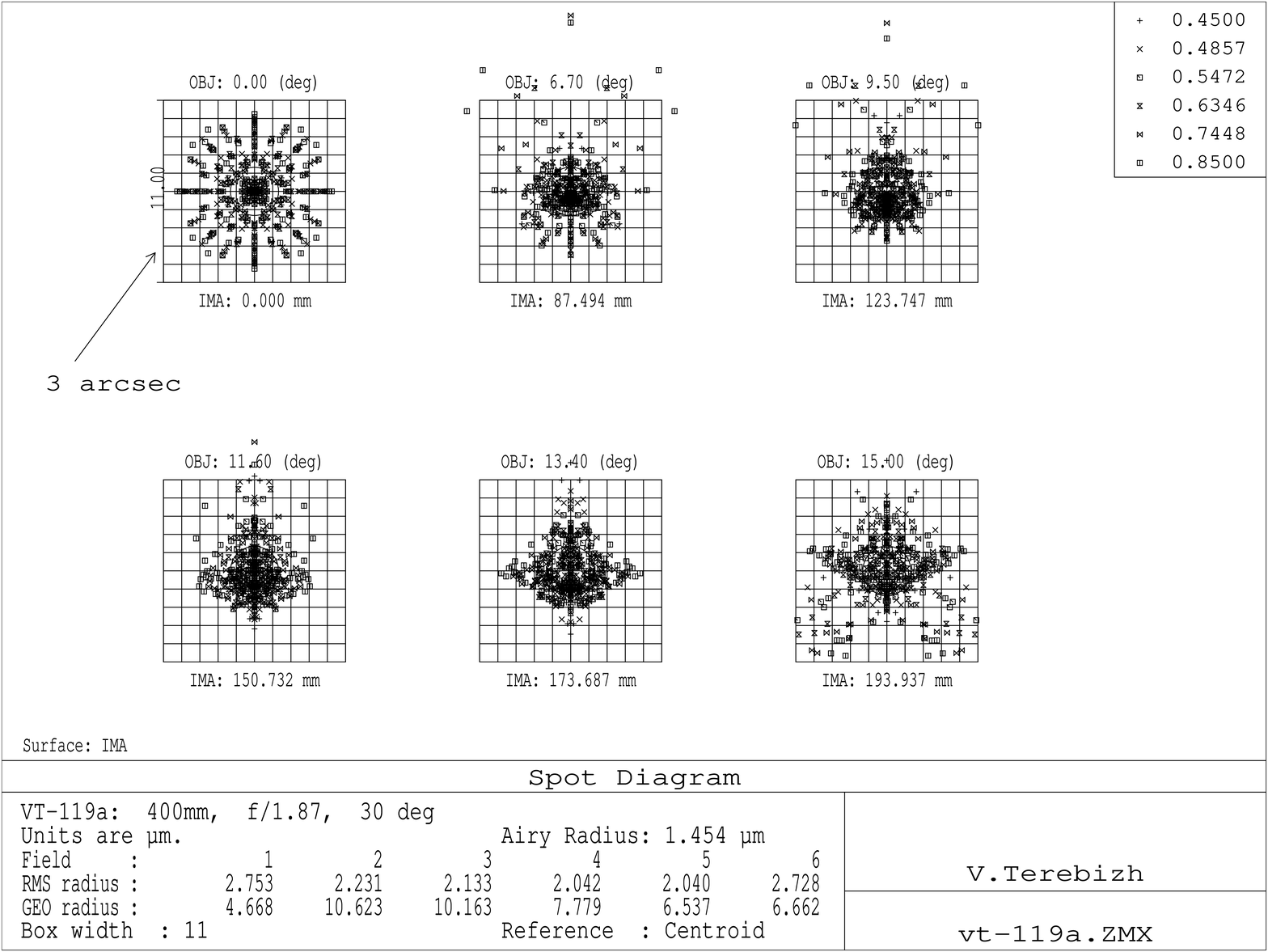} 
  \caption{Spot diagram of VT-119a in the polychromatic waveband $0.45-0.85\,\mu$m.
     The field angles are $0^\circ$, $6.7^\circ$, $9.5^\circ$, $11.6^\circ$, 
     $13.4^\circ$ and $15.0^\circ$ (equal areas). Airy disc diameter is 
     $2.9\,\mu$m, box width is $11\,\mu$m $\simeq 3''$ .}
\end{figure}

There are two main tasks in carrying out surveys of large areas of the sky:
1)~we need to cover {\em sequentially} the area in the shortest time; 
2)~the sky area we are interested in should be under {\em continuous} 
observation, as is the case when looking for the transient objects.  
To some extent, problems of the first kind can be solved with the help of 
wide-field telescopes with a flat field of view, which size is now 
reaching $10^\circ$. It is easily seen that problems of the second kind require 
too many flat-field telescopes, so it seems that the best way in this case 
is the creation of a single telescope with an extremely wide angular field, even 
at a spherical focal surface. Just this way was chosen in the second half
of the last century, when the Baker-Nunn and Maksutov-Sosnina cameras were 
put into operation. Their angular field attained tens of degrees, while 
shielding of light and curvature of focal surface were overcome by application 
of narrow emulsion tape. The main disadvantages of these cameras were the 
following: 1)~the lens surfaces were substantially aspheric;
2)~the demanding sorts of glass were used in correctors; 3)~nevertheless, the 
image quality was inadequate. For example, four surfaces of the Baker-Nunn 
camera were aspheres of 4th and 8th orders, the Schott KzFS2 and SK14 glasses
were applied, but the calculated image of a point-like source of light had 
nearly $100\,\mu$m ($40''$) diameter. 

It is quite surprising that an {\em all-spherical} system made of simplest 
types of glass yet provides image quality close to the diffraction-limited one 
in a wide integral waveband. An example shown in Fig.~1 has been designed for 
the aperture $400$~mm, effective focal length $750$~mm ($f/1.9$), waveband 
$0.45-0.85\,\mu$m and the field of view diameter $30^\circ$. All lenses can 
be made of the same material; we prefer to use the fused silica because of its
stability and excellent optical properties (in particular, high UV-transparency).
Fig.~2 shows that the image quality is nearly the same across the field. 
The diameter $D_{80}$ of a star image varies from $5.6\,\mu$m ($1.5''$) at 
the optical axis to $7.6\,\mu$m ($2.1''$) at the field edge. 
The complete description of the system is given in Table~1. 

The system described above proceeds from two generic versions of the Schmidt 
camera, which then developed by D.D.~Maksutov, D.G.~Hawkins, E.H.Linfoot, 
C.G.~Wynne and J.G.~Baker (see Terebizh 2011 for discussion and references).
Our goal was the complete elimination of the aspherical surfaces by entering
a double Wynne [1947] meniscus~-- the first and forth lenses; the two inner 
lenses were inserted to minimize the chromatic aberration. (However, a system 
with three lenses provides only slightly inferior images.) A similar 
double-meniscus corrector was applied by Baker in his Super-Schmidt design, 
but he placed inside a highly aspheric correction plate such as that introduced 
by Schmidt. 

\begin{table}
\caption{VT-119a design with an aperture of 400~mm and 
 $30^\circ$ field of view. $R_0$~-- paraxial curvature radius, 
 $T$~-- distance to the next surface, $D$~-- light diameter, 
 FS~-- fused silica. All surfaces are spheres.}
\label{symbols}
\begin{tabular}{@{}cccccc}
\hline
Surf.& Comments    & $R_0$       & $T$         & Glass   & $D$    \\
 No. &             & (mm)        & (mm)        &         & (mm)   \\
\hline
 1   & Lens 1      & 1166.622    & 47.500      & FS      & 532.3  \\
 2   &             & 1267.954    & 254.723     & --      & 510.4  \\
 3   & Lens 2 Stop & $-$1103.18  & 46.700      & FS      & 391.3  \\
 4   &             & $-$1002.28  & 220.793     & --      & 411.7  \\
 5   & Lens 3      & $-$498.974  & 48.600      & FS      & 500.0  \\
 6   &             & $-$439.472  & 39.973      & --      & 521.1  \\
 7   & Lens 4      & $-$446.110  & 70.000      & FS      & 535.2  \\
 8   &             & $-$635.820  & 1000.82     & --      & 595.1  \\
 9   & Primary     & $-$1555.69  & $-$803.901  & Mirror  & 1130.4 \\
10   & Image       & $-$733.005  & --          & --      & 388.0  \\
\hline
\end{tabular} 
\end{table}

The method of working with the spherical focal surface applicable currently is 
to use small flat detectors each of which equipped with a flattening lens. This 
way has been implemented, e.g., in the {\em Kepler} space telescope. In our 
example, the radius of curvature of the focal surface is 733~mm, so for a small 
flat detector of size, say, 25~mm the edge images are blurred up to $30$ microns. 
Our preliminary consideration shows that image quality can be improved 
considerably even by a single lenslet made of fused silica, and is fully 
recovered using the doublet of the same material. Of course, large curved light 
detectors, the production of which has just begun, will be used in future.

As regards losses of light, a detector in the form of a band of size 
$30^\circ \times 5^\circ$ ($39$~cm $\times 6.5$~cm) shields less than 7\% of 
flux. Evidently, the light detectors can be placed on the field freely, i.e., 
respectively the studied area of the sky. 

It may be expected that the instruments of this type will compete with large 
conventional telescopes both in sequential and in continuous sky surveys. 

I thank M.~Boer (Recherche CNRS ARTEMIS, France) for stimulating discussions.

\end{document}